\RequirePackage{fix-cm}
\documentclass[twocolumn]{svjour3}          
\smartqed  
\usepackage[numbers]{natbib}
\usepackage{graphicx}
\usepackage{hyperref} 
\hypersetup{
	colorlinks=true,	
	linkcolor=blue,		
	citecolor=blue,		
	filecolor=blue,		
	urlcolor=blue		
}
\newcommand{\sr}{S_{recov}}
\newcommand{\mr}{M_{recov}}
\newcommand{\lft}{Left }
\newcommand{\ryt}{Right }
\begin{document}

\title{Simulations of imaging extended sources using the GMRT and the U-GMRT
}
\subtitle{Implications to observing strategies}


\author{Deepak Kumar Deo \and
        Ruta Kale
}

\authorrunning{Deo and Kale} 

\institute{Deepak Kumar Deo \at
              National Centre for Radio Astrophysics, Tata Institute of Fundamental Research, Post Bag 3, Pune University Campus, Pune 411007, India\\
              Tel.: +91-9039748525\\
              \email{anudeo.5@gmail.com}           
           \and
           Ruta Kale \at
            National Centre for Radio Astrophysics, Tata Institute of Fundamental Research, Post Bag 3, Pune University Campus, Pune 411007, India\\
              Tel.: +91-20-25719234\\
              \email{ruta@ncra.tifr.res.in}           
}

\date{Received: date / Accepted: date}

\maketitle
\begin{abstract}
Astrophysical sources such as radio halos and 
relics in galaxy clusters, supernova remnants and radio galaxies have angular 
sizes from a few to several $10$s of arcminutes.
In radio interferometric imaging of such sources, the largest angular 
size of the source that can be imaged is limited by the  
shortest projected baseline towards the source. It is 
essential to determine the limitations of the recovery of the extended features 
 on various angular scales in order to interpret the radio image.
We simulated observations of a model extended source of 
Gaussian shape with the Giant Metrewave Radio Telescope (GMRT) using 
Common Astronomy Software Applications (CASA).
The recovery in flux density and in morphology of the model source was  
quantified in a variety of observing cases with changing source properties and 
the uv-coverage.    
If $\theta_{lar}$ is the largest angular scale sampled in an observation 
with the GMRT, then $>80\%$ recovery of a source of size 
$0.3\times\theta_{lar}$ is possible. 
The upgraded GMRT (U-GMRT) providing 200 MHz instantaneous bandwidth 
between 300 - 500 MHz will allow 
a factor of two better recovery of a source of size $\theta_{lar}$ as compared 
to the GMRT at 300 MHz with 33 MHz bandwidth. 
We provide quantitative estimates for the improvement in extended source 
recovery in observations at low elevations and long durations. 
The presented simulations can be carried out for future radio telescopes such 
as the Square Kilometre Array (SKA) for optimisation of observing strategies 
to image extended radio sources.

\end{abstract}

\section{Introduction}
\label{intro}
Imaging of astrophysical sources with extended features like supernova remnants, 
star-forming regions, radio galaxies and radio halos and relics in galaxy 
clusters is important as we can infer morphology, spectrum  and 
polarisation of the source. These can reveal the underlying physical 
mechanisms at work in these sources which in turn, would help us to understand 
the evolution of our universe in greater detail. 

Radio interferometers measure the Fourier transform of the sky at the spatial 
frequencies defined by the configuration of the telescope elements (antennas).
The quality of an image from a radio interferometer depends on the distribution 
of uv-coverage which is the set of all baseline vectors projected on a plane 
perpendicular to the source direction (called as uv-plane) during the 
observations. 
Fourier transform of the uv-coverage is 
called as the `dirty beam'. The radio interferometer data contains the sky 
distribution convolved with the dirty beam and deconvolution is used to get the 
actual sky distribution. The more filled and uniform is the uv-coverage, the 
better would be the shape of the dirty beam and hence a better image.

For a radio interferometer with N antennas, there are $^NC_2$ number of 
baselines. The uv-coverage will span between the shortest and the longest 
baselines ($B_{min}$ and  $B_{max}$) in a given observation towards a target 
and by earth rotation synthesis, each baseline vector will trace elliptical 
tracks in the uv-plane \citep{thompson}. The angular scales sampled in the sky 
distribution correspond to the spatial frequencies sampled in the uv-plane. The 
shortest length in the uv-plane will limit the largest angular scale sampled in 
the sky ($\sim \lambda/B_{min}$) and the longest baseline will limit the 
angular resolution ($\sim \lambda/B_{max}$) of the radio interferometer. 
Sources in the sky can have structures ranging from scales smaller than the 
resolution of the interferometers and larger than the largest sampled angular 
scale. Since the sky distribution is unknown, it is essential to determine the 
limitations of the recovery of the sky distribution on various angular scales 
in order to interpret the radio image.

The problem is severe in the cases of extended sources such as molecular 
clouds, 
nebulae and sources called the radio halos associated with galaxy clusters where 
the angular extents of the sources (e.g. several tens of arcminutes) are often 
larger than the largest angular scale sampled by the radio interferometer.
A one-dimensional (1D) simulation study was done by Wilner and Welch 
\citep{wilner} 
on quantifying the effect of holes in the uv-plane on the imaging of molecular 
clouds. They modelled a 1D Gaussian and uniform disk source distribution as a 
representation of extended molecular emissions and derived the expressions to 
quantify the effect of holes in their imaging. They found that 
the central brightness recovered from a Gaussian distribution characterised by 
FWHM $\sim$ $\lambda / B_{min}$ is only about $3\%$. They also found that the central 
hole in the uv-coverage gives rise to more significant image distortions than 
many scattered 
outer holes.

The Giant Metrewave Radio Telescope (GMRT), the Very Large Array (VLA), 
Westerbork Synthesis Radio Telescope (WSRT) and the Australia Telescope 
Compact Array (ATCA) are among the largest radio interferometers that are used 
extensively to observe the sky in radio frequency bands. These are radio 
interferometers with dish type elements. The VLA, WSRT and ATCA have one or 
more movable elements to increase the uv-coverage. The GMRT is a telescope with 
a fixed configuration with elements distributed in a hybrid configuration 
providing good uv-coverage at short and long baselines. It is among the few 
telescopes operating at frequencies $< 1$ GHz and has been extensively used to 
observe extended radio sources such as radio halos in galaxy clusters \citep[e.g][]{feretti2012,venturi2013}.
In case of surveys that target search of extended sources such as radio halos, 
putting upper limits on non-detections requires a procedure of injection of 
model radio halos and testing their recovery in the given dataset. This method 
has been successfully used in the GMRT Radio Halo Survey \citep[GRHS][]{venturi2k7,venturi2k8} and its 
extension, the EGRHS \citep{egrhs1,egrhs2}. The underlying issue is that the imaging of extended 
emission is highly sensitive to the uv-coverage in a given observation. 
Non-availability of some antennas in the array, corruption due to radio 
frequency interference can modify the uv-coverage significantly and lead poorer 
sensitivity to extended emission. 

The next generation instruments such as 
the Karl G. Jansky VLA (JVLA) and the upgraded GMRT (U-GMRT) are moving towards 
wide bandwidth receivers which will drastically improve the uv-coverage. A 
comparison of the new facilities with the old ones is necessary to be 
quantified to understand the strengths of the new systems.

In this study, we quantify the effect of uv-coverage on the 
recovery of the total flux density and morphology of an extended radio source 
using simulated GMRT observations of model extended sources. 
We studied the recovery as a function of the angular size of the source,  
position in the sky, observation duration for the GMRT. We also simulate 
observations with the U-GMRT and compare it with the GMRT.

The paper is organised as follows: Sec. \ref{simulations} gives an overview
of the simulation, Sec. \ref{results} presents the results which are discussed 
in Sec. \ref{discussion}. Summary and conclusions are presented in Sec. 
\ref{conclusion}.

\subsection{GMRT and U-GMRT}
\label{telescope}
\begin{figure*}
\includegraphics[scale=0.49]{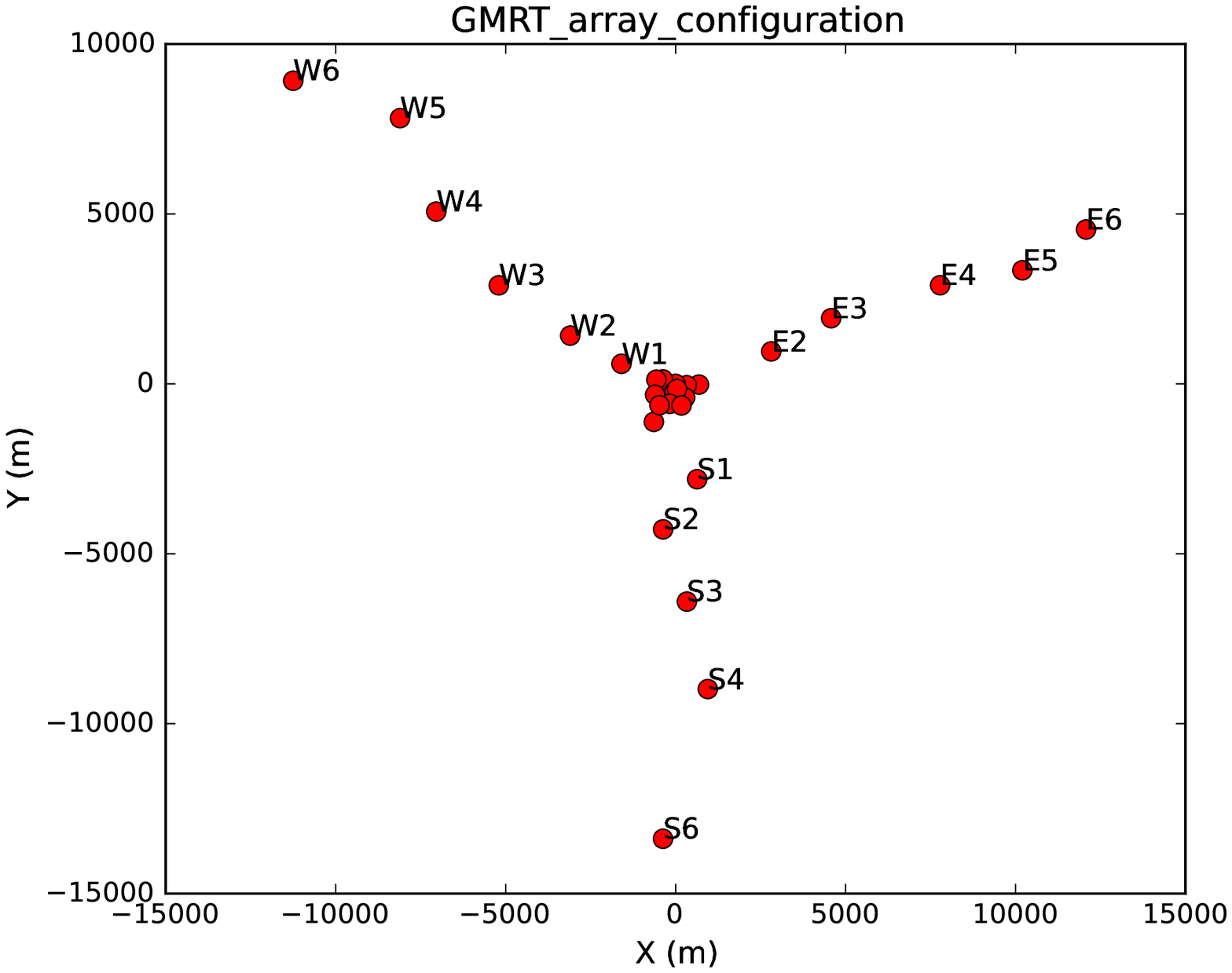}
\hspace{-1.0cm}
\includegraphics[scale=0.49]{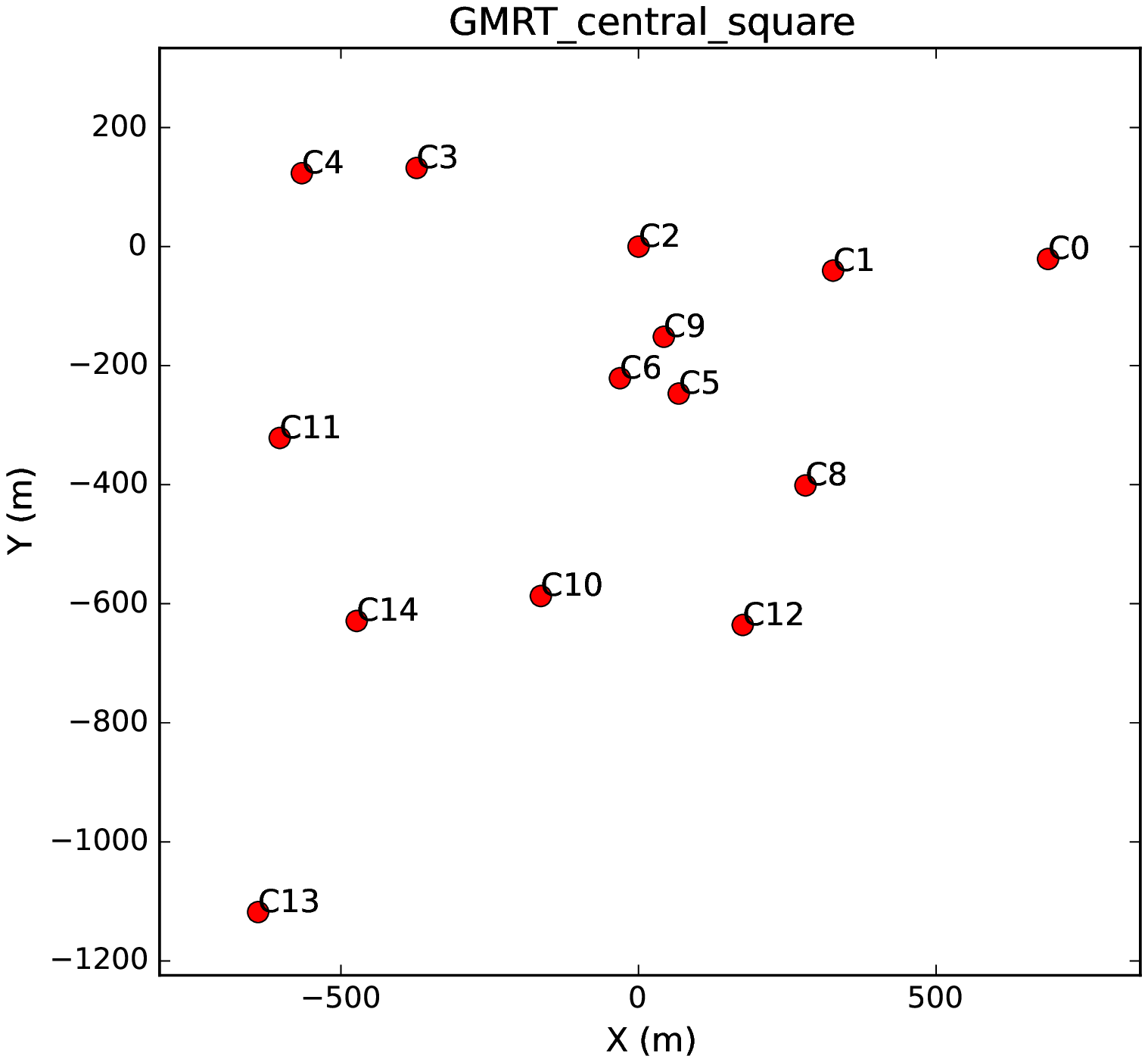}
\caption{\lft -- GMRT array configuration in local reference frame with the 
origin at the central square antenna C02. The arm antennas are labelled. Y axis 
is towards local north and X axis is towards local east. \ryt -- The central 
square containing 14 antennas is shown with the antenna labels.}
\label{gmrtArray}
\end{figure*}

Giant Metrewave Radio Telescope (GMRT) \citep{GMRT} is a radio interferometer 
set up by NCRA-TIFR at a site about 80 km north of Pune, India. GMRT 
(Latitude=$19.1^{\circ} N$, Longitude=$74.05^{\circ} E$) consists of thirty 45 m 
diameter antennas spread over a region of 25 km diameter. The antennas are 
distributed in a hybrid configuration, 14 of which are located in a central 
compact array of size $\sim$1.1 km and the rest 16 are distributed in a roughly 
`Y' shaped configuration (Fig. \ref{gmrtArray}), giving a maximum baseline 
length of $\sim$25 km. GMRT can observe in a declination range of $-53^{\circ}$ 
to $90^{\circ}$. It provides a maximum instantaneous bandwidth of 32 MHz for 
observation and can be used for observations between 130 MHz to 1450 MHz in the 
bands 130-170 MHz, 225-245 MHz, 300-360 MHz, 580-660 MHz and 1000-1450 MHz 
\citep{gmrt2,ugmrt2}. 
We have used the 610 MHz GMRT band as a fiducial band for simulations to 
quantify the limitations of recovering an extended radio source.

The upgraded GMRT (U-GMRT) will have new wide band receivers that will allow 
instantaneous bandwidth of 200 MHz in the frequency bands 120 - 250, 250 - 500 
and 550 - 850 MHz and of 400 MHz in the 1050 - 1450 MHz band \citep{ugmrt1,ugmrt2}. The 
broadband receivers will allow a drastic increase in the uv-coverage. In this 
work we compare the current GMRT and the U-GMRT in the 250 -500 MHz band in terms of 
capabilities to image extended radio sources. 

\section{Simulation overview}
\label{simulations}
The codes for the simulation of imaging model extended source 
and its observation with the GMRT were written using Python and 
the CASA tool-kit \citep{casa, casatoolkit}.
The following steps were carried out:
\begin{enumerate}
 \item A model sky containing a source of Gaussian shape was made. This was 
assumed to be the only source in the sky.
 The code was written to have the flexibility to allow choice of position in 
the sky, the width and strength of the Gaussian source.
\item The GMRT array configuration with the positions of the antennas in the local coordinate system with the antenna C02 as reference was used. The 
position of the reference antenna C02 on Earth was taken as listed in CASA.
\item The observing frequency, bandwidth, number of channels in the bandwidth, 
total observation duration and separation between scans were set according 
to the particular simulation experiment. 
\item The simulated observation was then imaged using the task `clean' in CASA (details in Sec.~\ref{clean})
\item The cleaned image was compared with the model image. 
\end{enumerate}

The aim of the simulations {\bf is} to quantify the recovery of extended 
source such as a radio halo when the uv-coverage is varied. The results 
of the morphology and flux density recovery varying the following parameters
are presented:
\begin{enumerate}
 \item source angular size,
 \item source flux density,
 \item observation duration, 
 \item source declination, and
 \item bandwidth.
\end{enumerate}
The case of ``source flux density'' variation is considered in order to 
demonstrate that the recovery does not depend on the source strength. Thus  
the results from the rest of the cases can be considered general enough to be 
applicable to source strengths other than that chosen here.

\subsection{Observing parameters}
\label{bands}
We made single channel visibilities of width 1 MHz as observed with the GMRT at 
610 MHz to study the recovery as a function of observing duration, source 
strength and position. For the bandwidth 
case, we simulated multichannel (each channel of 1 MHz) visibilities of 
bandwidth 33 MHz, 100 MHz and 200 MHz in the frequency 
band of 250-500 MHz.
Simulated observations with a bandwidth of 33 MHz represents GMRT whereas 
observations with a bandwidth of 100 MHz and 200 MHz represents observations with 
the U-GMRT. Integration time was taken to be 16 seconds in case of GMRT 
observation and 8 seconds in the case of the U-GMRT. 
The observation duration is two hours starting at the transit time of the source in the 
cases of angular size, source flux density and declination.
In the case of varying observing duration, multi-scan observations and bandwidth the 
start time of the observations is fixed to the rise time of the source.
 
\subsection{Source model}
\label{modelsky}
The motivation for our simulation study comes from the limitations of 
interferometers that are faced in imaging the extended sources towards galaxy 
clusters. Therefore we discuss the choice of our model based on the case of a 
radio halo though it is applicable to extended source of any origin.
Radio halos are Mpc sized diffuse sources associated with the intra-cluster 
medium of 
merging galaxy clusters \citep[e. g.][]{feretti2012} and are among the most 
challenging to image. 
We chose a 2D Gaussian as a simplistic representation of a radio halo. 
The linear size of our model halo was taken as 1 Mpc which corresponds to the FWHM of the 
chosen Gaussian and variation in its angular size represented the observation of 
radio halo 
at different redshifts.  For all the single 
channel studies, model image was made at 610 MHz. We 
referred to the radio halo in the galaxy cluster Abell 2163 \citep{feretti} for 
the realistic estimation of the total flux density ($S$) 
for our model source and found it to be 0.6 $Jy$. Model was also made over a 
range of redshifts ($z$) from 0.05 to 1.0 and its angular size ($\theta$) was 
calculated as a function of $z$ using the $\Lambda$-CDM 
cosmology \cite{parameters}. We chose $S$ for a model at z = 0.05 to be 0.6 $Jy$ 
and it was scaled according to the redshift. The $\theta$ 
and $S$ values of the model to the corresponding $z$ are given in Table 
\ref{tab:1}. The source size of 1 Mpc at the redshift of 0.05 corresponds to 
the largest angular size sampled at 610 MHz with the GMRT ($\lambda/B_{min}$).
Position of the model source was taken as (RA, Dec) $\equiv$ (4h, 
60$^{\circ}$) for all but the case where the declination was varied.

\begin{figure}
\centering
\includegraphics[scale=0.65]{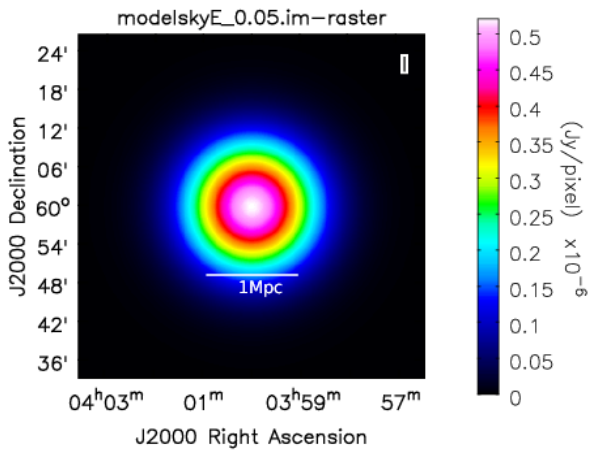}
\caption{2D Gaussian of width $16.8'$ at FWHM representing a radio 
halo of size 1 Mpc at the redshift of 0.05. Source is centred at position 
(4h,$60^{\circ}$). Total flux density ($S$) of the source is 0.6 $Jy$. 
The color scale shows the flux density in the units of $Jy$ pixel$^{-1}$.}
\label{model}
\end{figure}

\begin{table}[h!]
\centering
\caption{Angular size ($\theta$) and total flux density ($S$) of the source model 
at a particular redshift ($z$).}
\label{tab:1}       
\begin{tabular}{lll}
\hline\noalign{\smallskip}
$z$ & $\theta$ ($'$) & $S$ ($Jy$)  \\
\noalign{\smallskip}\hline\noalign{\smallskip}
0.05 & 16.8 & 0.6000 \\ 
0.10 & 9.0 & 0.1441 \\ 
0.13 & 7.3 & 0.0832 \\ 
0.15 & 6.4 & 0.0616 \\ 
0.17 & 5.9 & 0.0472 \\ 
0.2 & 5.2 & 0.0333 \\ 
0.3 & 3.9 & 0.0138 \\ 
0.4 & 3.3 & 0.0072 \\ 
0.5 & 2.9 & 0.0043 \\ 
0.6 & 2.7 & 0.0028 \\ 
0.8 & 2.4 & 0.0014 \\ 
0.9 & 2.3 & 0.0010 \\ 
1.0 & 2.2 & 0.0008 \\  
\noalign{\smallskip}\hline
\end{tabular}
\end{table}

\subsection{Using `clean' in CASA}
\label{clean}
CASA version 4.5.1 was used for analysis. Simulated visibilities were 
imaged using the task `clean'.
Since imaging of extended sources is involved, we used the option of 
multi-scale cleaning.
The multi-scale parameter in clean was set to have four values that were 
0, twice and ten times the synthesised beam and the fourth one equal to the 
angular size of the model source.
Image size was 
taken as twice the size of the primary beam of GMRT. 
Cell size was chosen to be 0.25 times the synthesised beam. 
Cleaning was done interactively until no unique peak could be identified 
visually  -- this is  when the residuals look uniform over the image. In this 
process an error is introduced based on where we stop the cleaning based on 
visual inspection (see Sec. \ref{diagnostics}).

\subsection{Diagnostics for comparison}
\label{diagnostics}
The recovery of the model image in the cleaned image was compared 
on the basis of the total flux density recovery ($\sr$) and the morphology 
recovery ($\mr$).
\paragraph{Flux density recovery}
The total flux density of the cleaned image 
($S_{cleaned}$) of the source was compared to that of the model ($S_{model}$) as:

\begin{equation}
S_{recov} \ \ = \ \ \frac{S_{cleaned}}{S_{model}} \times 100.
\end{equation}
The recovered flux density after `clean' has an error introduced due to 
the interactive cleaning. We examined the flux density 
and the value of peak in the residual images as measures of uncertainty 
on the recovered cleaned flux density. The residual flux density is 
$0.1 -  8\%$ and the peak residual flux density is $0.1 - 16\%$ of that in the 
model image. This can be considered as the typical uncertainty on the reported 
$S_{recov}$.

\begin{figure}
\begin{center}
\includegraphics[scale=0.49]{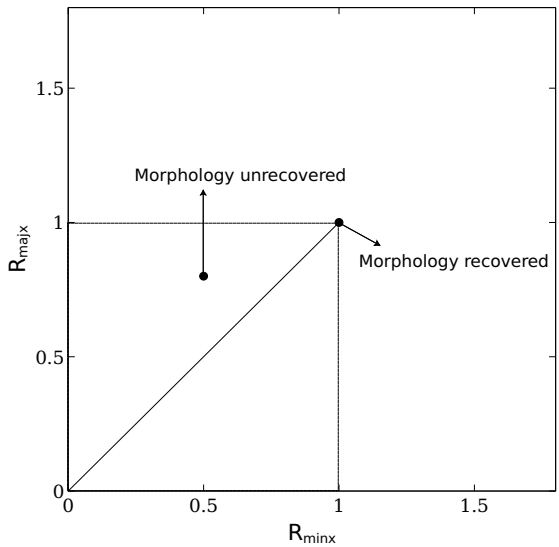}
\caption{Schematic representation of morphology recovery. A value at (1,1) 
 implies that the morphology of the source has been completely recovered and 
any deviation from it along the line $R_{majx}=R_{minx}$ or $R_{majx}>R_{minx}$ 
 implies incomplete recovery. No values will be seen in the region 
$R_{majx}<R_{minx}$ as by definition, the major axis is greater than the 
minor axis.}
\label{morpho}
\end{center}
\end{figure}

\paragraph{Morphology recovery}
The model image is a circularly symmetric 2D Gaussian. The departure from the 
circularity serves as a quantitative estimate of the recovered morphology. 
Gaussian fit was done over the cleaned and model images of the source to 
determine the shape (major and minor axes).
$\mr$ was quantified by comparing the ratio of the major axis of the cleaned image  
to that of the model ($R_{majx}$) with the ratio of their minor axes ($R_{minx}$) (Fig \ref{morpho}). 
The location of the point in the $R_{majx}-R_{minx}$ plane quantifies its departure from the 
model circular 2D Gaussian. An incomplete morphology recovery can also 
result in a value greater than $(1,1)$ in the plane when the recovered Gaussian 
is more extended with a slight decrease in its peak value. 
Error on the $\mr$ is as reported on Gaussian fits in CASA. This is 
between $0.01 - 0.1\%$ in all the cases and is not plotted in the figures.

\section{Results}
\label{results}
\subsection{Angular size}
\label{theta}
\begin{figure}
\centering
\includegraphics[scale=0.5]{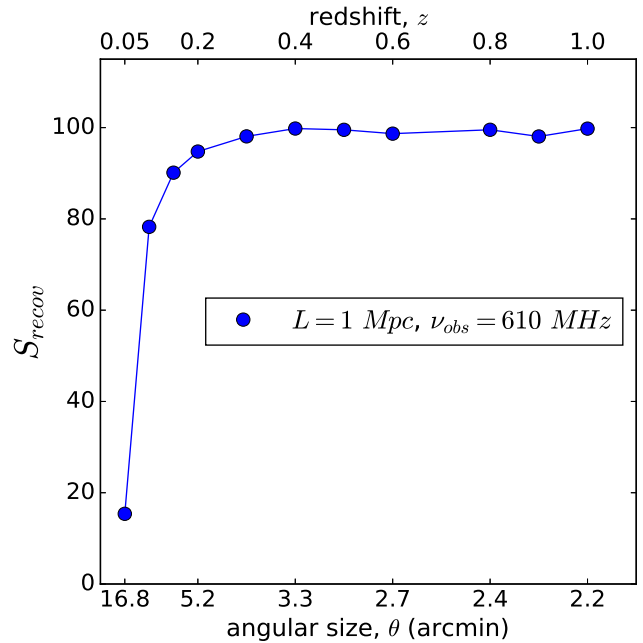}
\caption{$\sr$ as a function of $\theta$($z$) at 610 MHz GMRT 
observation. Source with $\theta = 16.8'$, largest $\theta$ that can be sampled at 610 MHz from GMRT, shows $\sr$ $\sim 14\%$. $\sr$ $\sim$ 100$\%$ for the sources with $\theta$ $<$ $5.2'$ or which are at $z > 0.2$.}
\label{rfVSz}
\end{figure}

The model source image of different angular sizes was made by 
varying $z$ from 0.05 to 1.0. In each case, an observation at 610 MHz 
for 2 hours starting at the source transit time was carried out. $S_{recov}$ $\sim$ 
$100\%$ was observed for the sources with $\theta$ $<$ $5.2'$ or which 
were at $z > 0.2$ (Fig. \ref{rfVSz}). Therefore in the remaining cases, we 
restrict to the angular size 
range $> 5.2'$.

\subsection{Source flux density}
\label{strength}
\begin{figure*}
\centering
\includegraphics[scale=0.49]{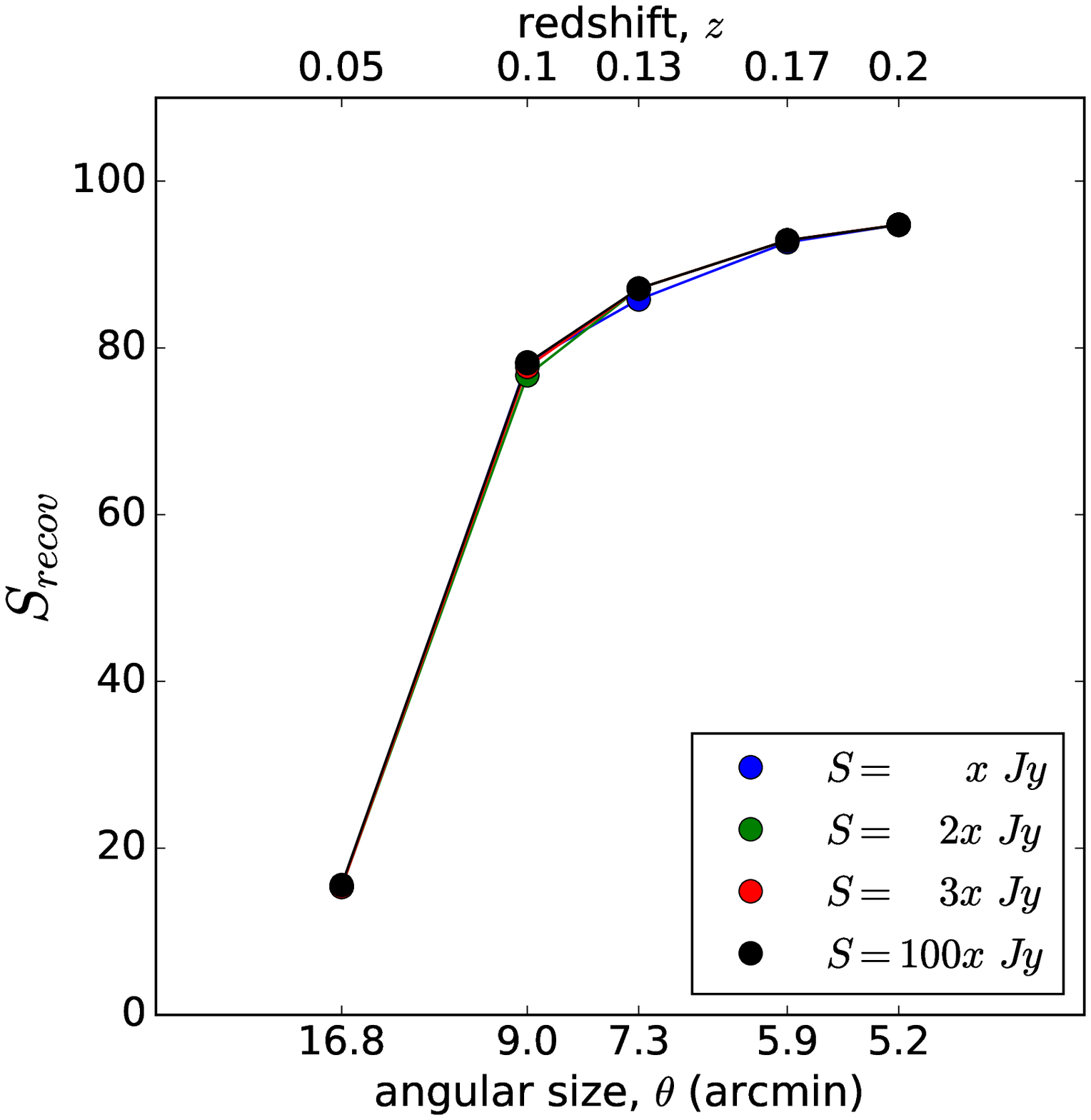}
\includegraphics[scale=0.49]{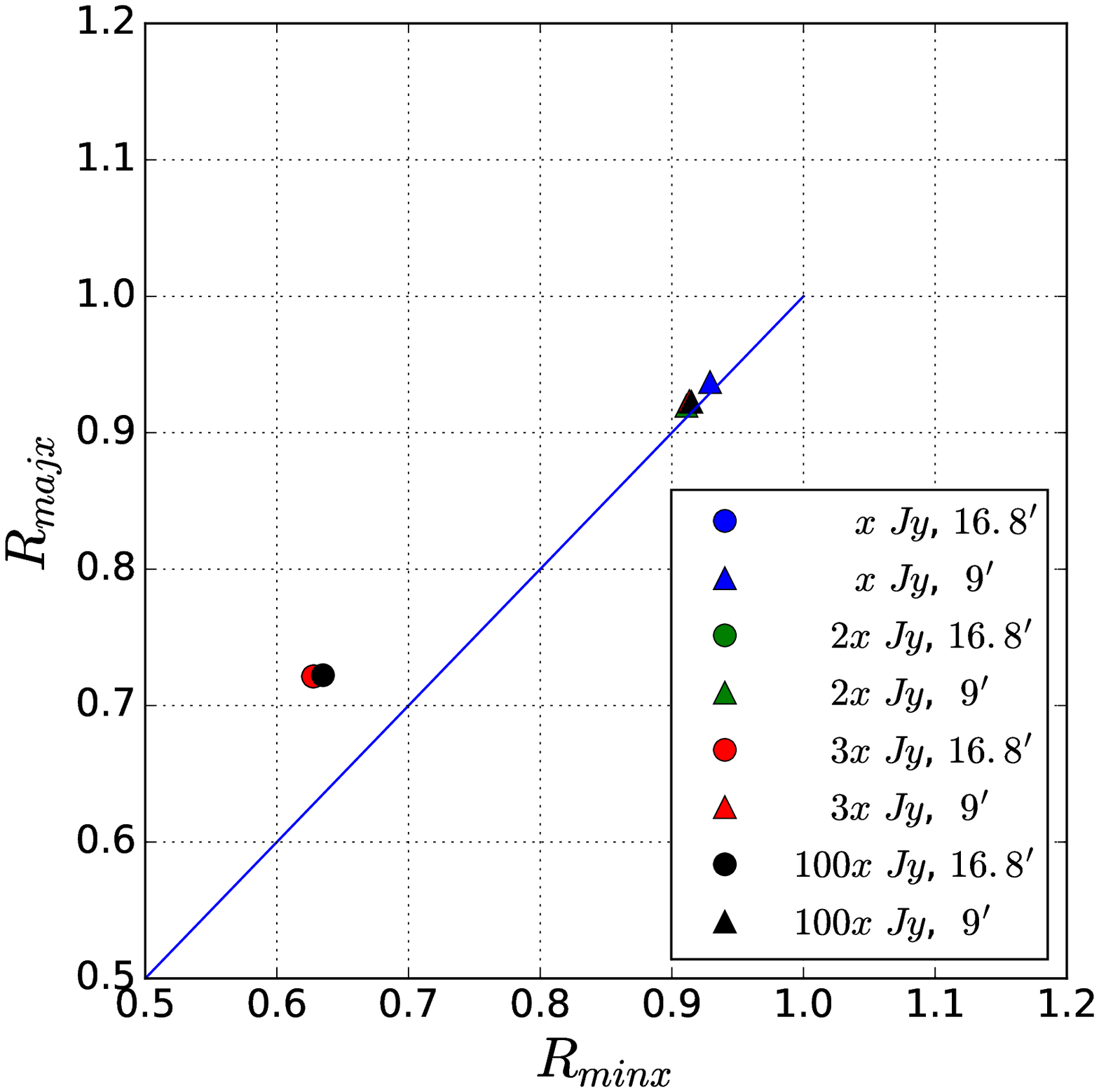}
\caption{Flux density case: \lft -- $\sr$ when total flux density ($S$) of the source is 
varied 
from $x=0.6$ Jy to $2x$, $3x$ and $100x$ Jy. $\sr$ doesn't vary with varying $S$. \ryt -- $\mr$ for the 
sources with $\theta = 16.8'$ and $9'$ when the source flux density is varied. 
$9'$ sources shows  $\sim 90\%$ recovery in morphology whereas $16.8'$ sources $73\%$ recovery in their major axis while $63\%$ recovery in their minor axis.}
\label{rmflux}
\end{figure*}

The strength of the model source scaled by factors of 
two, three and 100 times that of that at $z=0.05$ and kept rest of the 
observation parameters same as 
in the angular size case (section \ref{theta}).
The $\sr$ and $\mr$ of the source remains unchanged  with increase in source 
strength (Fig. \ref{rmflux}) basically due to the unchanged uv-coverage. Sources 
with $\theta = 16.8'$ shows $\sr$ of $14\%$ whereas $9'$ sources shows $\sim 
90\%$ recovery. For $9'$ sources, $\mr$ is $\sim 90\%$ whereas $16.8'$ sources 
shows $73\%$ recovery in their major axis while $63\%$ recovery in their minor 
axis.

\subsection{Observing duration}
\label{obs}
\begin{figure*}
\centering
\includegraphics[scale=0.49]{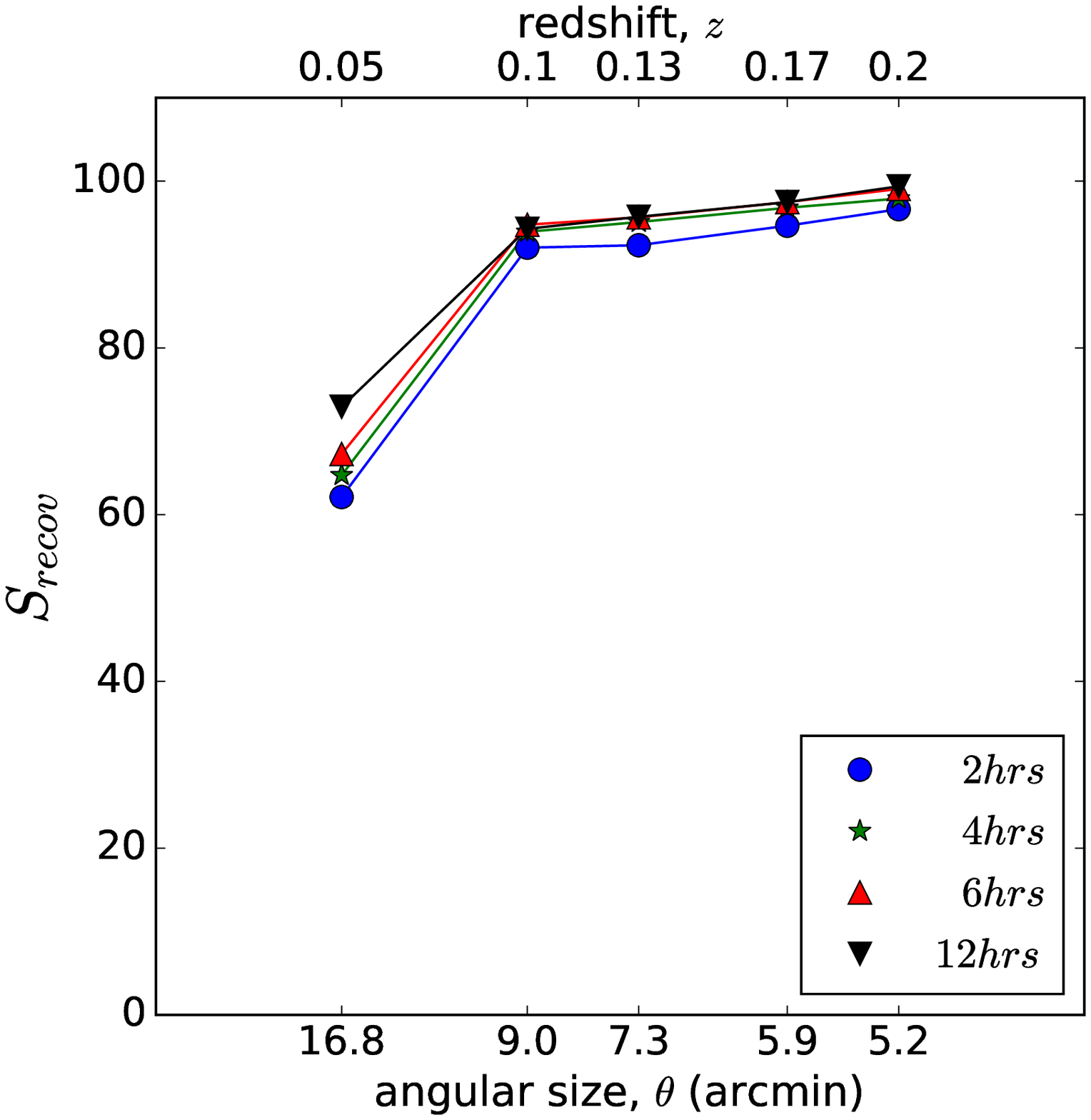}
\includegraphics[scale=0.49]{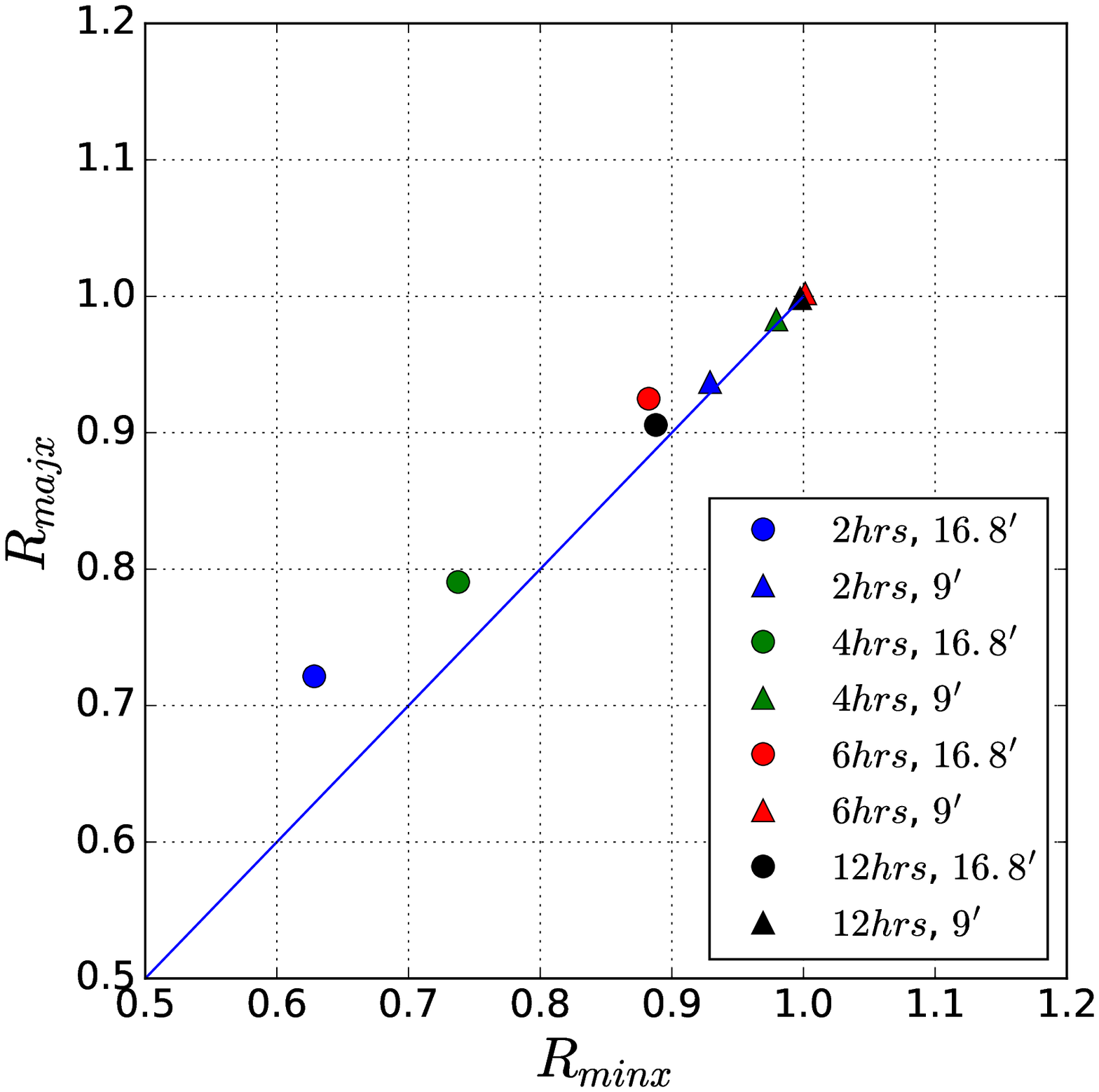}
\caption{Observation duration case: \lft -- $\sr$ when observation duration is varied. Source with $\theta = 
16.8'$ shows increase in its $\sr$ from $\sim 60\%$ in 2 hours to $73\%$ in 12 
hours observation. \ryt -- Comparison of $\mr$ when observation 
duration is varied. Source with $\theta = 
16.8'$ shows increase in its $\mr$ from $\sim 70\%$ in 2 hours to $90\%$ in 12 
hours observation. Source with $\theta = 9'$ shows $92\%$ $\mr$ in 2 hours and attains complete $100\%$ $\mr$ in 6 hours of observation.}
\label{rmobs}
\end{figure*}

The simulation was carried out with observing durations of 2, 4, 6 and 12 
hours, keeping the strength, position of the source\textbf{,} observing frequency and bandwidth 
constant. In all these cases the start time of the observation was chosen as the rise time of the source. The maximum time 12 hours, is the duration of full 
synthesis observation for the source at the chosen position. 
$\sr$  increases with the 
observation duration (Fig. \ref{rmobs}, \lft). 
For a source with $\theta = 16.8'$, $\sr$ increased from $\sim 60\%$ in 2 hours (from rise time) to $74\%$ in 12 hours observation. 
Significant improvement is seen in the case of $\mr$ (Fig. \ref{rmobs}, \ryt). For $16.8'$ source, $\mr$ increased from $\sim 70\%$ in 2 hours to $90\%$ in 12 hours. Source with $\theta = 9'$ showed $92\%$ recovery in 2 hours, $97\%$ in 4 hours and complete $100\%$ morphology recovery in 6 hours and 12 hours of observation. 

\subsubsection{Multi-scan observation}
\label{multiscans}
\begin{figure*}
\centering
\includegraphics[scale=0.49]{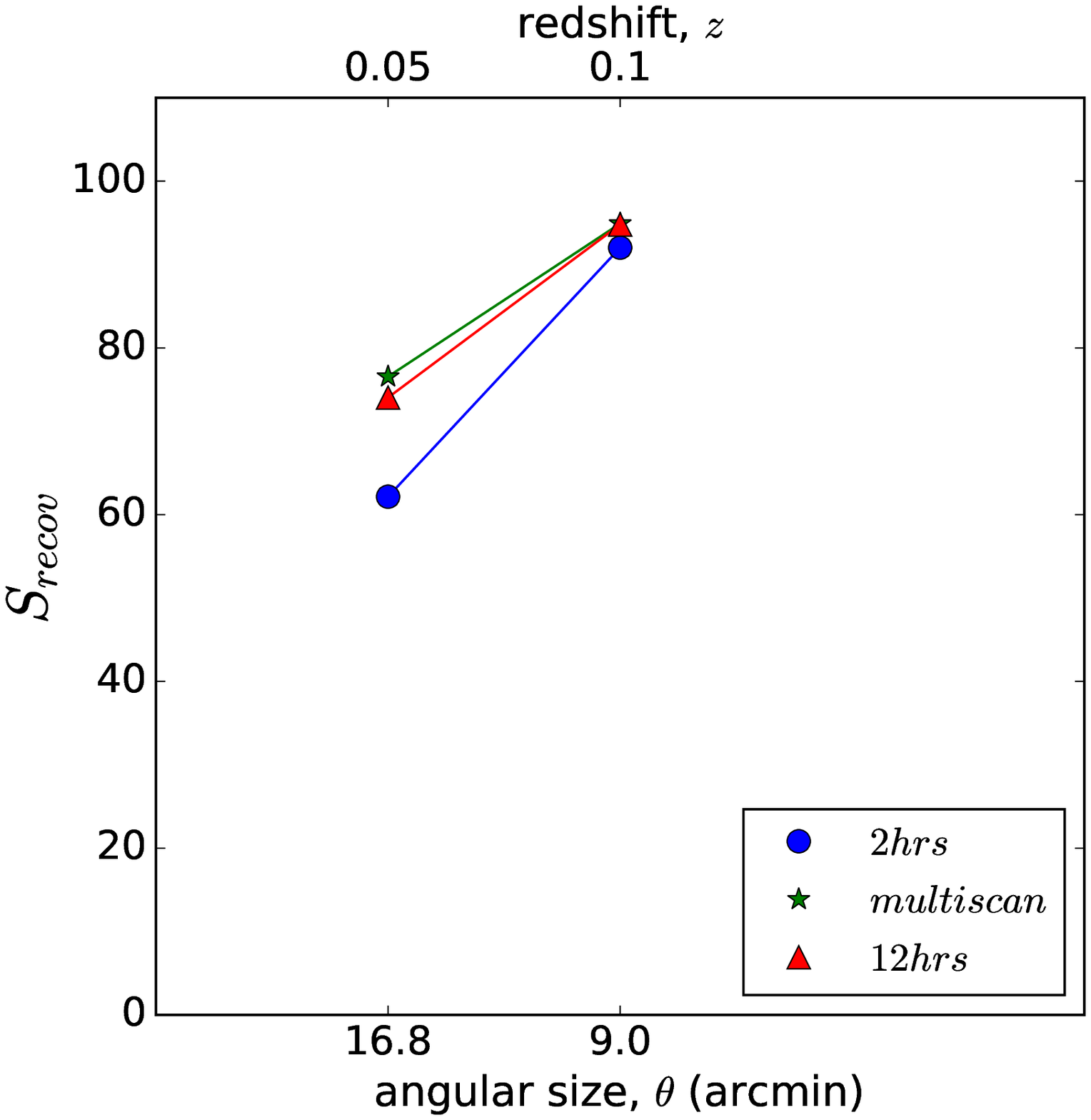}
\includegraphics[scale=0.49]{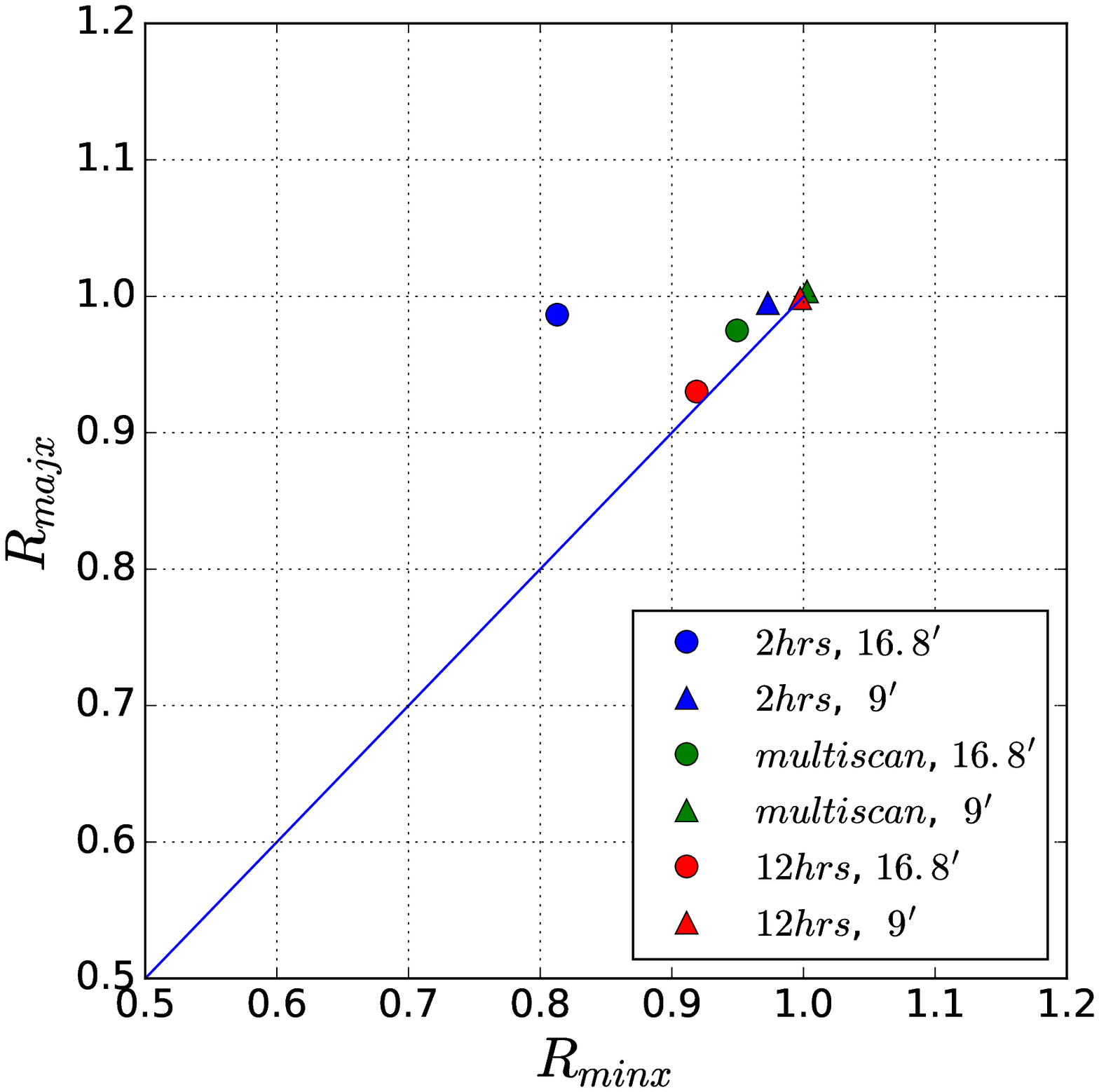}
\caption{Multi-scan case: \lft -- $\sr$ comparison of the multi-scan 2 hours 
observation to continuous 2 hours and 12 hours observation case. $\sr$ is similar for both multi-scan and 12 hours observation case. It is $\sim 77\%$ for $16.8'$ source and $92\%$ for $9'$ source.  
\ryt -- $\mr$ comparison of multi-scan 2 hours observation to 
continuous 2 hours and 12 hours observation case. $9'$ source shows $\sim 100\%$ recovery for all the 3 cases. $16.8'$ source shows above $90\%$ recovery in case of multi-scan and 12 hours observation whereas in case of 2 hours observation, $16.8'$ source shows $\sim 100\%$ recovery in its major axis but only $80\%$ in its minor axis thus attaining an elongated shape along its major axis.}
\label{rmsnaps}
\end{figure*}

A short total observing time spread in short scans over a longer duration is 
possible. This is a common strategy to maximise the uv-coverage in a given 
total time on the source. The 2 hours observation was spread in 6 discrete scans of 
20 minutes each at regular intervals spanning over 12 hours. Comparison was 
done only for the two largest angular size sources ($16.8'$ and 
$9'$). $\sr$ and $\mr$ are similar to the continuous 12 hours observation (Fig. 
\ref{rmsnaps}). $\sr$ for $\theta = 16.8'$ and $\theta = 9'$ is $\sim 74\%$ and $92\%$ respectively which is equivalent to the corresponding recoveries at 12 hours observation. $\mr$ is above $90\%$ for multi-scan 2 hours case and 12 hours case for $16.8'$ source, $9'$ source shows complete recovery for both the cases.  

\subsection{Declination}
\label{dec}
\begin{figure*}
\centering
\includegraphics[scale=0.49]{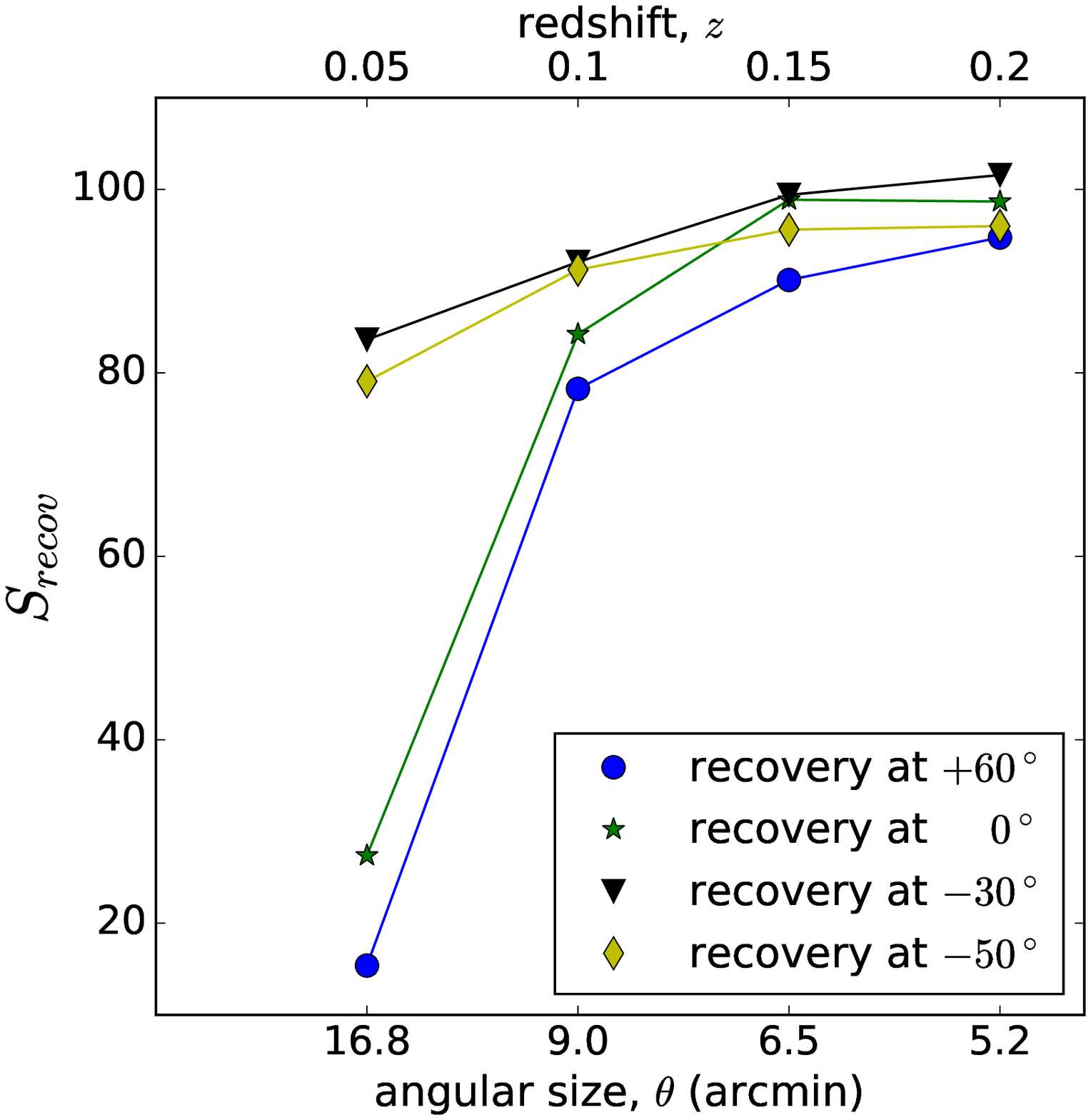}
\includegraphics[scale=0.49]{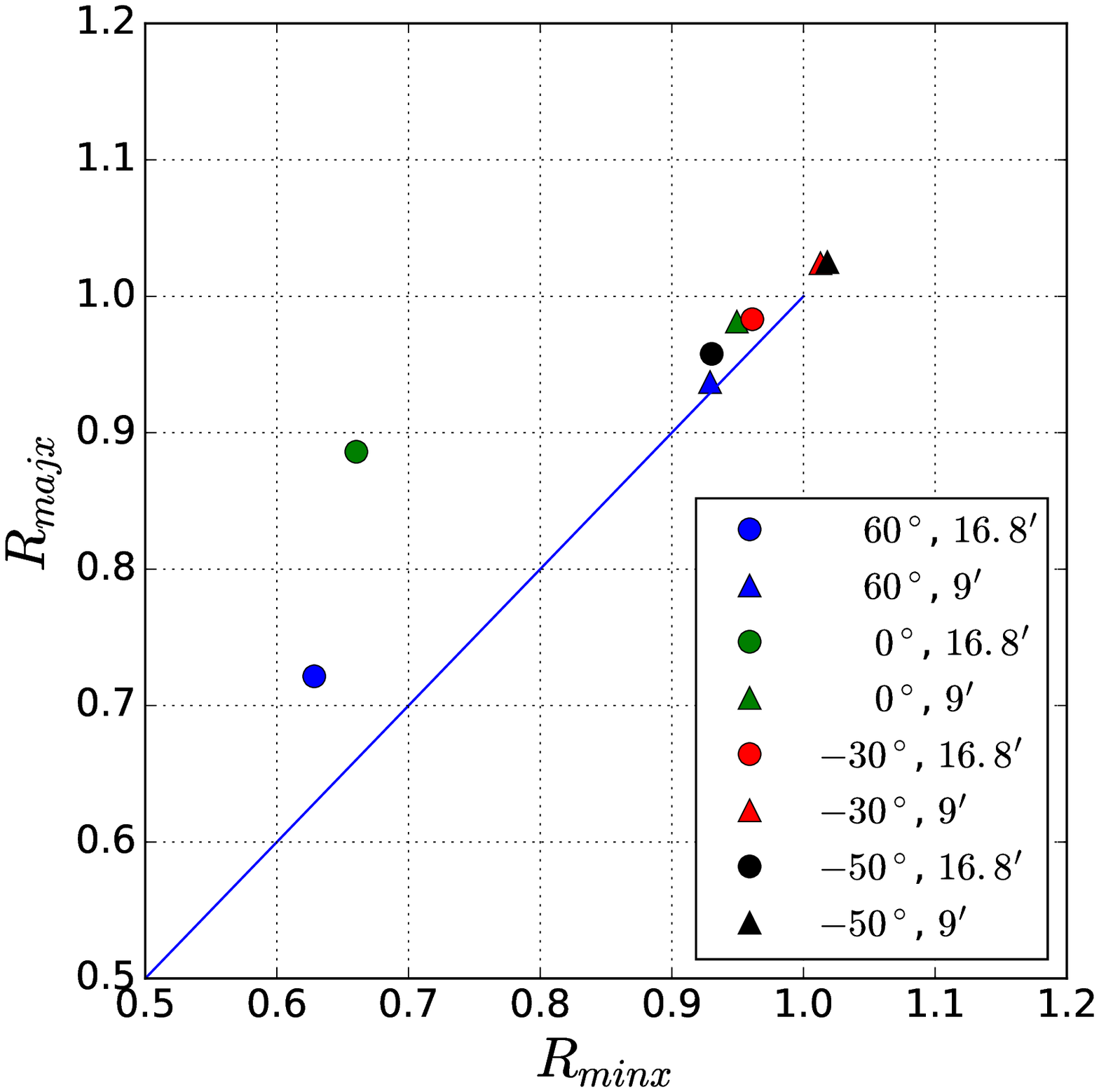}
\caption{Declination case: \lft -- $\sr$ when source declination is varied. 
$\sr$ is $<30\%$ for source with $\theta=16.8'$ at $0^\circ$ and $60^\circ$ and it is $\sim 80\%$ for the same source at $-30^\circ$ and $-50^\circ$. 
\ryt -- $\mr$ when source declination is varied. For all declinations, $9'$ source shows $\mr$ $>90\%$. $16.8'$ source shows $\mr$ $>90\%$ at $-30^\circ$ and $-50^\circ$. At $60^\circ$ $\&$ $0^\circ$, it recovers its shape by $\sim 70\%$ $\&$ $\sim 90\%$ along its major axis respectively and $\sim 60\%$ $\&$ $\sim 65\%$ along its minor axis respectively.}
\label{rmdec}
\end{figure*}

The distribution of baseline lengths changes as a function of declination due 
to projection. 
 We chose declinations $60^{\circ}$, $0^{\circ}$, $-30^{\circ}$ and 
$-50^{\circ}$ that span the observing declination range of GMRT and 
studied the recoveries for a 2 hours observation from transit. 
 At lower declinations (below $-30^{\circ}$), we have better $\sr$ (Fig. 
\ref{rmdec}, \lft) and $\mr$ (Fig. \ref{rmdec}, \ryt). For source with $\theta = 16.8'$, $\sr$ at $0^\circ$ and $60^\circ$ is $< 30\%$ whereas it is $\sim 80\%$ below $-30^\circ$. For all the declinations, source with $\theta = 9'$ shows $>90\%$ $\mr$. In case of $16.8'$ source, $\mr$ is $>90\%$ at $-30^\circ$ and $-50^\circ$. At $60^\circ$, it recovers its shape by $\sim 70\%$  and $\sim 60\%$ along its major axis and minor axis respectively. At $0^\circ$, it shows $\mr$ of $\sim 90\%$ and $\sim 65\%$ along its major axis and minor axis respectively.

\subsection{Bandwidth: GMRT and U-GMRT}
\label{bandwidth}
\begin{figure*}
\centering
\includegraphics[scale=0.49]{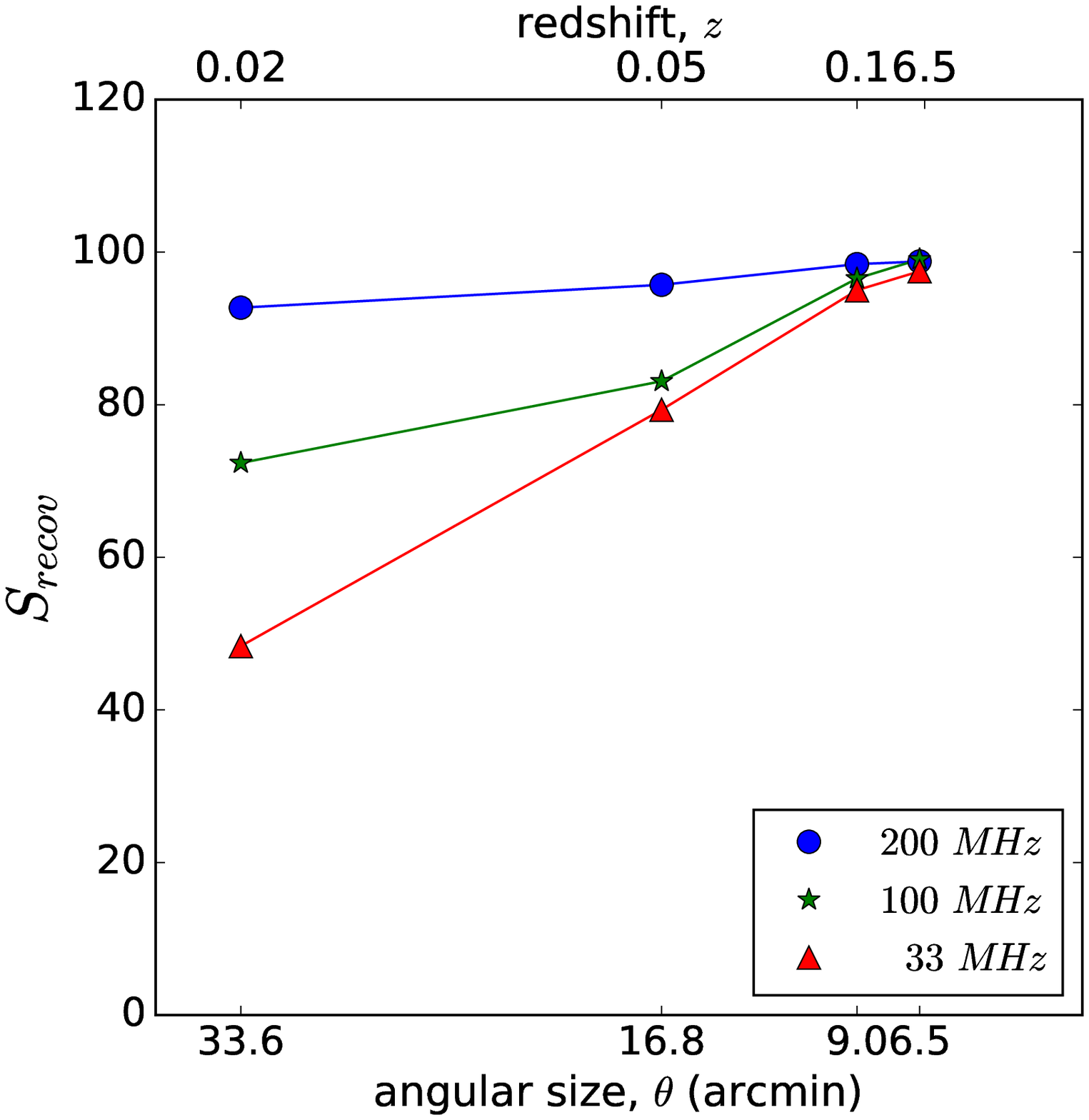}
\includegraphics[scale=0.49]{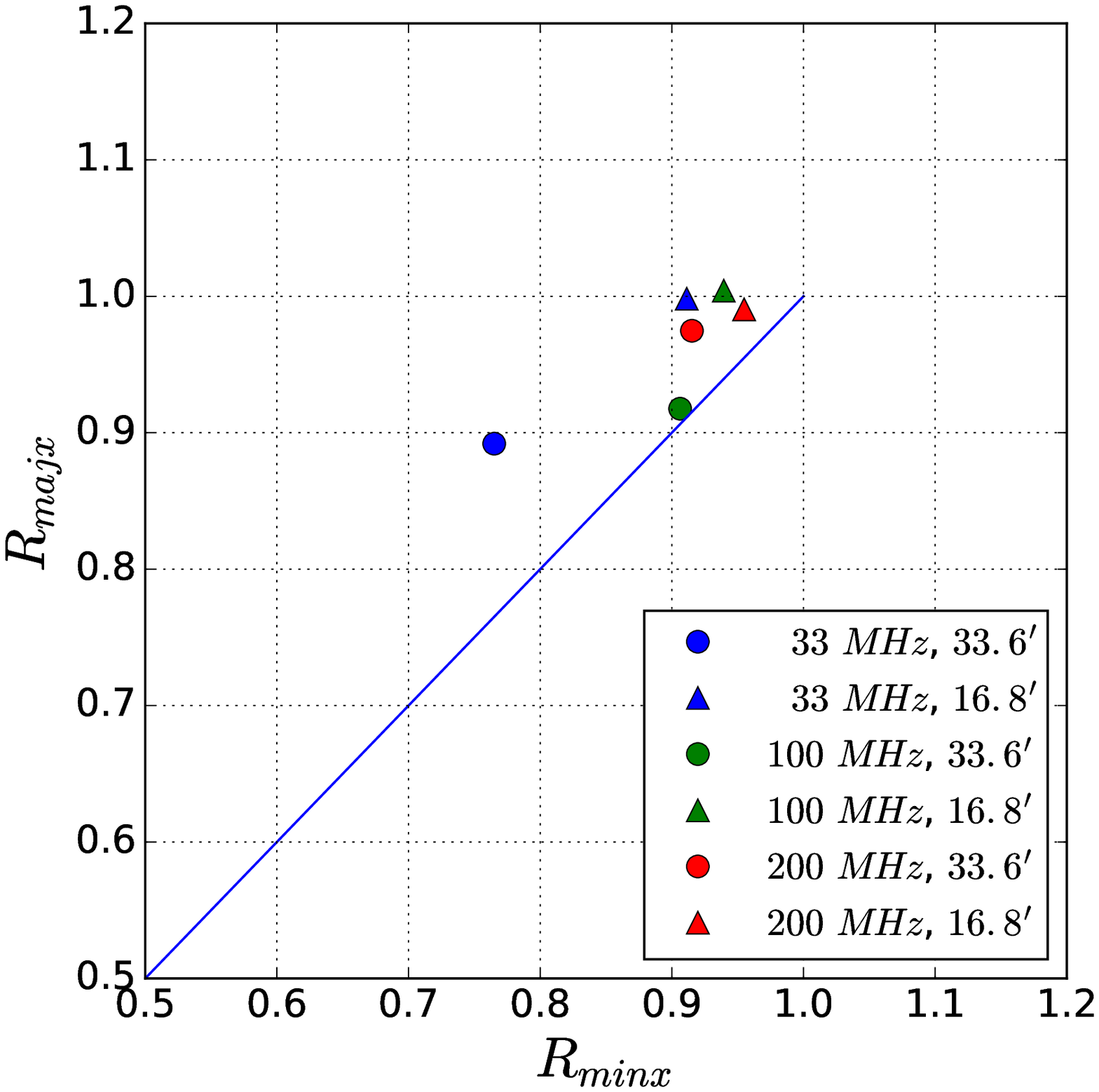}
\caption{Bandwidth case: \lft -- $S_{recov}$ when observation bandwidth is varied. $\sr$ in case of 200 MHz bandwidth (U-GMRT) increases by a factor of 2 as compared to that of at 33 MHz bandwidth (GMRT) while 
observing $33.6'$ source which is the largest $\theta$ that can be sampled at 
300 MHz from GMRT. \ryt -- $M_{recov}$ when observation bandwidth is 
varied. Sources observed at U-GMRT bandwidth (100 $\&$ 200 MHz) shows $\mr$ $>90\%$.}
\label{rmBW}
\end{figure*}

We made multichannel observations having instantaneous bandwidths of 
33 MHz representing the GMRT, 100 MHz and 200 MHz 
representing the U-GMRT. The starting frequency in all these cases was fixed 
to 300 MHz. Observations were carried out for 2 hours from rise of the source. Increase in bandwidth (BW) 
increases the $\sr$ significantly and U-GMRT (BW=200 MHz) shows $\sim 90\%$ flux density
recovery as compared to that of $45\%$ from GMRT even for the source with the largest angular 
size ($\theta = 33.6'$) that can be sampled at 300 MHz (Fig. \ref{rmBW}, \lft). Increasing bandwidth also improves the 
recovery in morphology. At 200 MHz, source with $\theta = 33.6'$ shows $98\%$ recovery 
in major axis as compared to that of $90\%$ at 33 MHz, recovery in minor axis 
increases from $76\%$ to $92\%$. For $16.8'$ source, GMRT and U-GMRT both shows 
 $\mr$ $>90\%$ (Fig. \ref{rmBW}, \ryt). 

\section{Discussion}
\label{discussion}

Imaging of the sky with radio interferometers is a challenge due to the fact 
that the measurements are limited to the sampled angular scales and the sky 
distribution is an unknown. Therefore simulations of observations and imaging 
of extended sources are important to understand the limitations of the 
interferometer. 

In this work, we presented the simulations of observations of a 2D Gaussian 
source with the GMRT at 610 MHz. Observations were simulated with varying  
source declination, angular size, observing duration and bandwidth. The 33 MHz 
GMRT observation was compared to the U-GMRT 200 MHz bandwidth observation.
The changes in the  parameters reflected the change in the uv-coverage. The 
recovery of 
the total flux density and morphology of the source was quantified in each of 
the cases. The observed visibilities were chosen to be free of random noise 
in order to isolate the effect of the uv-coverage on the recovery of the model 
source after ``cleaning''.

\subsection{A rule of thumb for extended source recovery}
The simulations were carried out at the fiducial frequency of 610 MHz. 
The results obtained in the case of changing source declination, observation 
duration and source angular size can be scaled for interpretation at any other 
frequency. A recovery of $>90\%$ was obtained for sources of angular size 
$5.2'$ at 610 MHz in 2 hours. The ratio $5.2'/16.8' = 0.3$ can be used to find 
the angular size at other frequency. For example, at 300 MHz the largest 
angular scale sampled with the GMRT is $33.6'$. The angular size at which 
the source will be fully recovered is $0.3\times33.6' = 10.1'$, which can be 
also seen to be true from the simulation for the 
33 MHz bandwidth case shown in Fig.~\ref{rmBW} (left). Therefore if 
$\theta_{lar}$ 
is the largest angular scale sampled with the GMRT at a given frequency, then 
near complete recovery sources of angular size, $\theta_{recov} = 
0.3\times\theta_{lar}$ is possible. This can be considered as a thumb 
rule for estimating the approximate scale at which a recovery above $80\%$ is 
possible in a short observation ($\sim$ 2 hours).

The simulations presented here are applicable in general to the recovery 
of extended sources with an interferometer with uv-coverage similar to that of 
the GMRT. We illustrate the use of our results with a particular 
case of surveys that target search of extended sources of 1 Mpc size in galaxy 
clusters. The redshift where the angular sizes correspond to the size of 1 
Mpc are plotted in Figures 4 - 9. If a target is located at declination 
$\delta_{src}$ at a redshift of $z_{src}$ and is planned to be observed for a 
duration of $t_{src}$ with a bandwidth $\Delta \nu$, then the recovery of a 1 
Mpc source of Gaussian type profile can be located in the plots. This can serve 
as a first step to check the feasibility of the survey in a given redshift 
range for objects of certain sizes with Gaussian profiles. 

\subsection{Recommendations for observing strategies}
\label{implications}
\begin{figure}[h!]
\centering
\begin{center}
\includegraphics[scale=0.5]{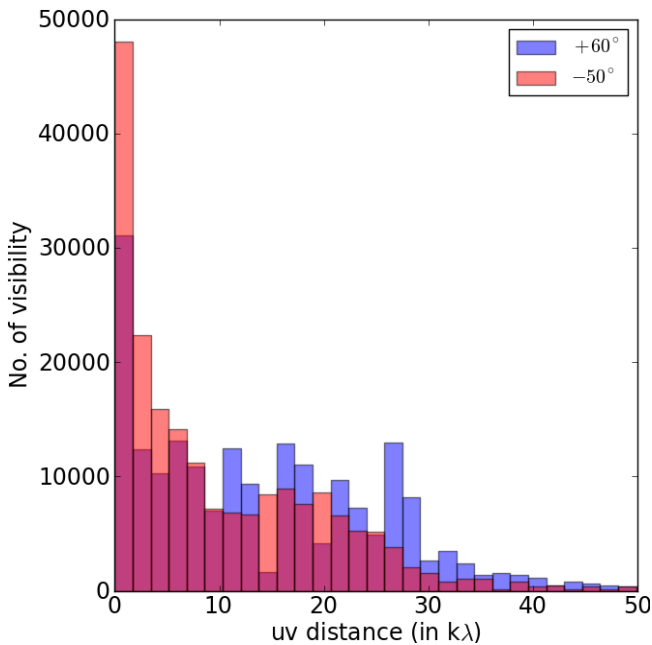}
\caption{Visibility distribution comparison of the source at 610 MHz observation from GMRT at the declinations of 
$60^{\circ}$ and $-50^{\circ}$. Dark pink region is common to both 
the histograms.}
\label{histdec}
\end{center}
\end{figure}

The results of this simulation can be used in deciding the strategies for 
targeted surveys for the search of diffuse radio emission of large angular 
scales. 
The recommendations for observing strategies are as follows:

\begin{itemize}
 \item Cycling through sources to improve uv-coverage: From the simulation of 
 a 2 hours observation divided into multiple scans spread over twelve hours it 
 was shown that the improvement in uv-coverage provides a recovery of extended 
source nearly as good as a twelve hour observation (Sec. \ref{multiscans}, Fig. \ref{rmsnaps}). This strategy is often used 
when there are multiple targets in the sky that can be cycled through. An 
example of this strategy is the Australia Telescope Low-Brightness 
Survey \citep{2010MNRAS.402.2792S}.
\item Low elevation observations: The shortest projected baselines of an array 
are when the targets are observed at the lowest possible elevation. The density 
of short baselines goes up as one observes a target at low elevation. For the 
GMRT the number of visibilities within 0 - 5$k\lambda$ can be a factor of 1.5 
higher for targets with declinations $-50^{\circ}$ as compared to those at 
declinations $60^{\circ}$ (Fig. \ref{histdec}). The low declination sources represent targets 
observed at low elevations for the GMRT. If a 
target is observed for a limited duration, then in order to have the best 
chance of recovering all the extended emission, the observation be done  
either at the rise or setting time of the source when the lowest elevation data 
can be recorded. Observing at low declinations can be disfavoured for other 
reasons. At the GMRT one of the concerns with low elevation observing is 
the higher level of radio frequency interference coming in through terrestrial 
sources.
\item Long observations: Full synthesis observations are needed in order to 
recover the morphology of the source. A full synthesis observation is 
recommended if the angular size of the source to be recovered is close to that 
of the largest angular scale sampled by the interferometer.
\item Wide bandwidths: U-GMRT is going to be a wide band system offering 
instantaneous bandwidths of 200 - 400 MHz over fixed frequency bands in the 
range 120 - 1500 MHz. Wider bandwidths imply radially broad tracks in the 
uv-plane resulting in better uv-coverage. For a model
source with $\theta = 33.6'$ we recovered $\sim 90\%$ of the source flux density 
from 200 MHz bandwidth observation as compared to $70\%$ from the 100 MHz 
bandwidth and $45\%$ from the 33 MHz bandwidth observation.
\end{itemize}

\section{Conclusions}
\label{conclusion}
We have simulated the observations of extended sources from GMRT and U-GMRT 
and quantified their limitations in the recovery in flux density and morphology 
of the sources when various parameters affecting the uv-coverage were varied. 
Results from this study are summarised below:
\begin{itemize}
 \item $\sr$ is $\sim 100\%$ from GMRT for sources with $\theta < 5.2'$ when 
observed for 2 hours from transit at 610 MHz.
\item $\sr$ improves with increasing observation duration, however, 
improvement is insignificant due to the noise-free simulation of the 
observations. Improvement is more significant in \textbf{the} case of morphology recovery as 
compared to the flux density recovery. A 2 hours observation when spread over 
12 hours (full synthesis) observation gives the similar recovery ($\sr$ $\&$ 
$\mr$) as full synthesis observation.
\item Low declination observation $<-30^{\circ}$ show near $80\%$ 
recovery as compared to the higher declinations due to shorter projected 
baselines. 
\item $\sr$ of the source with largest angular size that can be sampled in P-band (300 - 500 MHz) is better by a factor of two from U-GMRT as compared to the GMRT. 
\end{itemize}
Based on the simulations, as a thumb rule, in a GMRT observation at a given 
frequency, $>80\%$ recovery of a source of angular size, $\theta_{recov} = 
0.3\times\theta_{lar}$ is possible.

The simulations have quantified the recovery of extended sources 
with the GMRT and U-GMRT and can be used to plan observations. Such simulations 
can be carried out using the same codes for any interferometer such as 
the JVLA and the Square Kilometer Array (SKA).

\begin{acknowledgements}
We thank the referee for the valuable comments. RK acknowledges the support through the DST-INSPIRE Faculty award. DKD is supported 
through the DST-INSPIRE project fund. We thank N. Mohan for insightful discussions. We thank D. Oberoi for providing the 
GMRT antenna coordinates. GMRT is run by the National Centre for Radio 
Astrophysics of the Tata Institute of Fundamental 
Research.
\end{acknowledgements}

\bibliographystyle{spmpsci}      
\bibliography{casasim}   

\end{document}